# Half-metallicity in Armchair Boron Nitride Nanoribbons: A First-Principles Study

Hari Mohan Rai[1], Shailendra K. Saxena[1], Vikash Mishra[1], Ravikiran Late[1], Rajesh Kumar[1], Pankaj R. Sagdeo[1], Neeraj K. Jaiswal[2] and Pankaj Srivastava[*3]

[1]Material Research Lab. (MRL), Indian Institute of Technology Indore, Indore (M.P.) – 452017, India
[2]PDPM- Indian Institute of Information Technology, Design and Manufacturing, Jabalpur – 482005, India
[3]Computational Nanoscience and Technology Lab. (CNTL),
ABV- Indian Institute of Information Technology and Management, Gwalior – 474015, India
* Corresponding author: pankajs@iiitm.ac.in

### ABSRTACT

Using density functional theory, we predict half-metallicity in edge hydrogenated armchair boron nitride nanoribbons (ABNNRs). The predicted spin polarization is analyzed in detail by calculating electronic and magnetic properties of these hydrogenated ABNNRs by means of first-principles calculations within the local spin-density approximation (LSDA). ABNNRs with only edge B atoms passivated by H atoms are found to be half-metallic (regardless of their width) with a half-metal gap of 0.26 eV. Upto 100% spin polarized charge transport is predicted across the Fermi level owing to the giant spin splitting. Transmission spectrum analysis also confirms the separation of spin up and spindown electronic channels. It is revealed that H-passivation of only edge N atoms transforms non-magnetic bare ribbons into energetically stable magnetic semiconductors whereas hydrogenation of both the edges does not affect the electronic and magnetic state of bare ribbons significantly. The results are promising towards the realization of inorganic spintronic devices.

**Keywords:** Boron nitride; Nanoribbons; Bloch states; Electronic structure; D. Density of states.

INTRODUCTION

Half-metallicity is the key property for spin transport based electronic devices. It has been widely studied and predicted in different nanoscale materials [1–5] accompanied with other compounds like double layer perovskites, [6],[7] alloys (including ternary and quaternary Heusler alloys)[8–10], transition metal pnictides, and chalcogenides in zinc blende phase.[11–13] Recently, an exceptionally large room temperature spin polarization of $(93^{+8}_{-11})$% has been directly evidenced in an epitaxial thin film of $Co_2MnSi$.[14] Present article predicts half-metallicity/spin polarization in armchair boron nitride nanoribbons (ABNNRs) via edge hydrogenation. Quasi-one 1-D thin strips carved out of hexagonal boron nitride (h-BN) sheet(s), [5], [15–17] properly known as BNNRs, are of high research interest because, as compared to their parent counterpart(s), they also exhibit excellent structural stability and distinct electronic properties due to the quantum confinement effects. [5] Synthesis of hollow BNNRs using B–N–O–Fe as precursor has already been reported by Chen. et al..[18] Similar to their organic counterparts, Graphene Nanoribbons (GNRs), BNNRs can also be fabricated from a single h-BN layer via lithographic patterning.[19] Depending on the definite shape of edges BNNRs are of two type - armchair boron nitride nanoribbons (ABNNRs) and zigzag boron nitride nanoribbons (ZBNNRs). It has been already proven that BNNRs exhibit very fascinating electronic properties similar to that of GNRs, [20,21] which are very important from the application point of view. The band gap ($E_g$) of BNNRs whose armchair edges are passivated by hydrogen oscillates periodically with increasing width [22] whereas the band gap of ZBNNRs decreases monotonically when both the edges are H-passivated. [23,24] It has been already shown through first principles calculations that spin-polarization may be realized also in BNNRs either with the application of an external in-plane electric field [25] or by chemically functionalizing zigzag edges with different groups such as H and F.[26], [27] In our previous study on ZBNNRs, [23] we predicted that one-edge (only B-edge) H-termination makes the ribbons 'semi-metallic' in nature unlike 'half metallic' as predicted by Zheng et al., [26] but for spin polarized transport both of these results are in agreement with each other. Furthermore, half-metallicity has also been revealed theoretically in ZBNNRs through structural modifications (e.g., by making stirrup, boat, twist-boat etc.) [28] and percentage hydrogenation.[29] In case of ribbons with armchair edges, spin polarization has been predicted in hybrid armchair BCN- nanoribbons.[30] However, any systematic and detailed study for analyzing the effect of edge hydrogenation on electronic, magnetic and transport properties of ABNNRs has not been reported so far.

Here, we predict half-metallicity in ABNNRs, induced due to the H-passivation of only edge B atoms. In addition, this article presents detailed and systematic analysis of electronic, magnetic and transport properties for edge hydrogenated (i.e. ribbon edges are passivated partially/fully with H atoms) ABNNRs of different widths. Present analysis reveals that H-passivation of only edge N atoms ($ABNNR_{HN}$) transforms bare ribbons into energetically stable 'magnetic semiconductors' unlike 'nonmagnetic semiconductors' as predicted by Ding et al..[31] On the other side, H-termination of only edge B atoms ($ABNNR_{HB}$), regardless of ribbon width, converts bare ABNNRs into ferromagnetic/anti ferromagnetic half-metals.

## COMPUTATIONAL DETAILS

The first-principles calculations were performed with Atomistix Tool Kit-Virtual NanoLab (ATK-VNL). [32,33] we employed ATK-VNL simulation package which is based on density functional theory (DFT) coupled with non-equilibrium Green's function (NEGF) formalism. The exchange correlation energy was approximated by local spin density approximation (LSDA) as proposed by Perdew and Zunger. [34] The reason for selecting LDA is that the generalized gradient approximation underestimates the surface–impurity interactions. [35] The ABNNRs were modeled with periodic boundary conditions along *z*-axis, whereas the other two dimensions were confined. The energy cutoff value of 100 Rydberg was selected for the expansion of plane waves. We implemented double $\zeta$ plus polarized basis set for all the calculations. The *k*-point sampling was set to $1\times1\times100$. In order to avoid artificial inter-ribbon interactions, ribbons were separated using a cell padding vacuum region of 10 Å. All the atoms, in the considered geometries, are relaxed and optimization of atomic positions and lattice parameters has been continued until the forces on each constituent atom reduced upto 0.05 eV/Å. We represent the ribbon-width by a width parameter $N_a$, defined as the number of B/N atoms along the ribbon width as depicted in Figure 1, therefore, ABNNR with n B/N atoms across the ribbon is named as *n*-ABNNR.

## RESULTS AND DISCUSSION

For present calculations we have considered ABNNRs with – (i) fully H-passivated edges (ABNNR$_{HBN}$) and (ii) half H-passivated edges. Furthermore, the ABNNRs belonging to later category are divided into two subgroups i.e., ABNNR$_{HB}$ and ABNNR$_{HN}$ depending upon whether only B or only N edge atoms are passivated by H-atoms respectively. The electronic and magnetic properties of bare ABNNRs have also been investigated and said ribbons were found to be nonmagnetic semiconductors in consistent with Ding et. al. [31]. In order to take size effects into considerations, we investigate ABNNRs having widths $N_a$ = 6 to 10. Since the findings are qualitatively similar for symmetric ribbons (Odd $N_a$), we displayed the figures only for ABNNRs with $N_a$ = 9. Figure 1 schematically represents optimized geometry of 9-ABNNR in all considered ribbon configurations with $N_a$ = 9 as representative case. The corresponding convention of super cell, used for simulation, is also depicted for various ribbon structures.

The calculated electronic band structures with corresponding density of states (DOS) for 9-ABNNR$_{HB}$ and 9-ABNNR$_{HN}$ are illustrated in Figures 2 and 3 respectively. A significant difference in the form of giant splitting of spin states can be clearly observed in, both, band structure as well as DOS profile. For 9-ABNNR$_{HB}$, the spindown channels [Figure 2 (a)] are metallic as shown with red lines, α [valence band maximum (VBM)] and β [conduction band minimum (CBM)] crossing the Fermi level. On the other hand, due to the presence of a larger band gap (4.64 eV) the spin-up (blue) bands exhibit insulating behavior as shown in Figure 2 (b). This half-metallic behavior is qualitatively independent of width and is observed for all investigated ABNNR$_{HB}$ structures. We predict that the charge transport across the Fermi level through all ABNNR$_{HB}$ structures should be dominated by the spin-down electrons. The same behavior is also reflecting in the corresponding DOS profile [Figure 2 (c)] where the crossing of Fermi level by spindown states is observed. The transmission spectrum (TS), [20] plotted as transmission coefficient (TC) versus energy graph in the inset of Figure 2 (b), also confirms this half-metallic nature of ABNNR$_{HB}$. A small peak (TC = 3.1), observed in spindown TS across the Fermi level, indicates the presence of conducting channels for spindown electrons whereas an energy gap (i.e., separation between spin up transmission states across the Fermi level) of about 4.6 eV points towards the unavailability of conducting channels for spin up electrons. The absence of mirror symmetry between up and downspin spectra is correlated with the splitting of spin states as observed in band structure and DOS of ABNNR$_{HB}$ structures. Thus, it is expected that the current flow though such systems should be entirely spin polarized. The half-metal gap of 0.26 eV has been observed for ABNNR$_{HB}$ which is large enough for room temperature operation and also comparable to the values reported for half-metallic ZBNNRs and GNRs.[26,36] The splitting of spin states is witnessed also in ABNNR$_{HN}$ structures [Figure 3 (a) and (b)]. However, unlike ABNNR$_{HB}$, a semiconducting/wide gap semiconducting behavior has been observed for all tested ABNNR$_{HN}$ structures with a band gap of ~2.75 eV (for spindown bands). The TS, shown in the inset of Figure 3(a), exhibits complete absence of conduction channels around Fermi level as the value of TC is zero across it, which also points towards the existence of a wide band gap in ABNNR$_{HN}$ structures.

Figure 4 (a) and (b) depicts the electronic band structure and corresponding DOS for ABNNR$_{HBN}$ ($N_a$=9, as a representative). It is obvious from figures that ABNNR$_{HBN}$ structures are insulating ($E_g$> 4 eV) in nature with degenerate spindown and spin-up bands. The DOS and TS (shown in corresponding insets) also support this insulating behavior. Further, the observed mirror symmetry in the peaks of opposite spin-states, reveals the spin degeneracy of electronic states. Moreover, the width dependence of band gap shows an oscillating behavior for ABNNR$_{HBN}$ as depicted in the inset (2) is a characteristic signature of ABNNRs.[22] In order to understand the origin of half-metallicity in ABNNR$_{HB}$ structures we have estimated the bare on site Coulomb repulsion[27,37] as a difference between ionization potential and electron affinity [27,38] for each element individually. The estimated values are 12.88 eV for H, 8.02 eV for B and 14.79 eV for N. This large difference ($\geq$ 4.86 eV) between N/H and B atoms inhibits the charge transfer through edge-B atoms/B-H bond; therefore, the electrons are localized only in 2$p$ orbital of N atoms. This mechanism for origin of half-metallicity is consistent with present PDOS analysis and previous reports [27,39] on BNNRs.

For the further confirmation of charge localization in unpassivated edge-N atoms of ABNNR$_{HB}$, Bloch states have been calculated as shown (for $N_a$=9 as a representative) in Figure 5. The structure of 9-ABNNR$_{HB}$ [Figure 1(a)] consists of 27 B, 27 N and 6 H atoms. A single B (N) atom contributes three (five) valence electrons as its electronic configuration is 2s$^2$2p$^1$ (2s$^2$2p$^3$) whereas a single H atom has only one valence electron- 2s$^1$. Thus there are total 222 valence electrons in the system. Moreover, each band is doubly degenerated (considering spin) and hence there will be a total of 111 valence bands. Therefore, the band index, which we used in calculations, for VBM is 110 (α band) and 111 (β band) for CBM as indexing starts from zero. For convenience, the spindown band, existing just below (above) the α (β) band [Figure 2 (a)], is designated as (α-1) [(β+1)] band. The band index of spindown (α-1) and (β+1) bands are 109 and 112 respectively whereas the band index for spin up α' and β' bands [Figure 2 (b)] are 108 and 113 respectively. The presented Bloch states calculations are performed for k-points value of (0.0, 0.0, 0.5). In order to keep the size of figures in presentable format, the contour and isosurface of 9-ABNNR$_{HB}$ for α, β, (α-1), (β+1), α' and β' bands are shown, only with one third of the unit cell, in Figure 5 (a) and (b) respectively. For a clear view, the results of α band are also presented with complete unit cell which is used for calculations. It is clear that the Bloch states (yellow and blue orbitals) corresponding to the (α-1), (β+1), α' and β' bands are completely distinguishable as no mixing of blue or yellow orbital is present in corresponding isosurface [Figure 5 (b)]. ABNNR$_{BHN}$ structures also exhibit similar results (as shown in figure S1 in the supplementary information) as no band crosses the Fermi level [Figure 4 (a)]. The localization of electronic charge across the un-passivated edge N atoms, as observed in the corresponding contour, confirms that the dangling bond of edge N-atoms is mainly responsible for observed half-metallic behavior of ABNNR$_{HB}$ structures. Moreover, the contour and isosurface (for α and β bands particularly), shown in Figure 6 (a) and (b) respectively, are almost identical (considering charge localization) as both of these bands correspond to the same spin states (downspin) across the Fermi level. It can be clearly observed from contour and isosurface that the unpaired electrons are mainly localized at the bare edge-N atoms. In addition, the interchanged colors of Bloch states for α and β bands, at top ribbon edge (as pointed by solid black arrows), indicates separation of these states via Fermi level which is in consistent with the separation of (α-1), (β+1), α' and β' bands (as indicated by dotted red arrows). Conversely, colors, at the bottom ribbon edge (as pointed by solid green arrows), are same for α and β bands which points towards the encroachment of α band (β band) into CB (VB). A little charge localization is also appearing around H and near edge N atoms but it has no contribution across the Fermi level as confirmed via PDOS analysis. In addition, the distance between any two adjacent edge N atoms (~2.43 Å) is large enough to ensure that no edge reconstruction has taken place, and there is consequently only one dangling bond per edge-N atom is present. The present analysis suggests that the half-metallic spin polarization in ABNNR$_{HB}$ structures is entirely attributed to the localization of unpaired electrons at bare edge-N atoms.

We also assess the effect of different kinds of edge H-passivation on magnetic properties of bare ABNNRs. The magnetic moment M (µ$_B$) and magnetic stabilization i.e. the energy difference ΔE (eV) between spin polarized (E$_{LSDA}$) and spin compensated states (E$_{LDA}$) for most stable (lowest Cohesive Energy) edge-hydrogenated ribbon configuration (ABNNR$_{HN}$) are presented in Table I. A noteworthy net magnetic moment of about 2.02 µ$_B$ (2.04 µ$_B$) is witnessed for all investigated ABNNR$_{HN}$ (ABNNR$_{HB}$) structures. On the contrary, all considered ABNNR$_{HBN}$ structures do not possess any magnetism. Present non-$d$ orbital type of magnetism observed in ABNNR$_{HB}$ structures is accounted for the magnetic moment of unpaired electrons localized in bare edge-N atoms. This is consistent with giant splitting of spin states as shown in Figure 2 (a-b) and localization of electrons as discussed in Bloch state analysis. It is also clear from Table I that, for ABNNR$_{HN}$, the spin polarized state is ~0.55 eV more stable with respect to the spin un-polarized state (E$_{LDA}$). Moreover, the coupling between electrons of inter-edge N atoms may be either ferromagnetic or antiferromagnetic, as the observed energies are very close to each other. In summary, we reveal that the electronic and magnetic properties of ABNNRs critically depend on the type of edge hydrogenation, as summarized in Table II.

Finally, in order to examine, which of the observed results are likely to be possible to achieve experimentally, the relative stability of considered ribbon configurations, has been calculated in terms of cohesive energy ($E_c$) by using the following relation;

$$E_C = \frac{1}{(x+y+z)}\left[E_{tot}^{ABNNR} - \left(xE_{atom}^B + yE_{atom}^N + zE_{atom}^H\right)\right] \quad (1)$$

where $E_{tot}^{ABNNR}$, $E_{atom}^{B}$, $E_{atom}^{N}$ and $E_{atom}^{H}$ are the total energies of ABNNR, isolated B, N and H atom respectively; $x$, $y$ and $z$ are the number of B, N and H atoms in the ribbon structures under investigation. The $E_c$ calculated for considered ribbon structures are summarized in Table III. Smaller is the value of $E_c$, more stable is the structure. Therefore, it is revealed from Table III that the stability is increasing with ribbon width for all the considered configurations. This indicates that amount of stress in the nanoribbons causing structural instability, is inversely proportional to the ribbon width. Since, the electronic and magnetic properties of the BNNRs depend critically on the type of edge passivations, (i.e., type of edge hydrogenation for present case), the understanding of different types of edge hydrogenations is crucial. The perusal of Table III indicates that irrespective of their width ABNNR$_{HN}$ structures are energetically most stable amongst all three edge H-passivated configurations under investigation. ABNNR$_{HB}$ and ABNNR$_{HBN}$ structures are 0.07 eV and 0.26 eV less stable as compared to ABNNR$_{HN}$.

## CONCLUSION

Conclusively, the effect of edge hydrogenation on the electronic and magnetic properties of ABNNRs has been systematically investigated using ab-initio calculations. ABNNR$_{HN}$ structures, regardless of ribbon width, are found to be magnetic semiconductor and energetically most favorable. The spin polarized states for all investigated ABNNR$_{HN}$ configurations are energetically more stable by ~0.55 eV as compared to spin compensated state. We revealed that Passivation of only edge-B (N) atoms via H atom i.e. ABNNR$_{HB}$ (ABNNR$_{HN}$) transforms non-magnetic insulating ABNNRs into magnetic half-metals (wide band gap semiconductor). It also causes giant splitting of spin electronic states which gives rise to a substantial net magnetic moment of about 2.04 (2.02) $\mu_B$. Conversely, both-edge H-passivation of bare ABNNRs does not alter their electronic or magnetic behavior significantly; therefore, ABNNR$_{HBN}$ structures are non-magnetic insulators with degenerate/symmetric spin states similar to bare ABNNRs. Energetically, half-metallic ABNNR$_{HB}$ (ABNNR$_{HBN}$) structures are 0.07 eV (0.26 eV) less stable as compared with ABNNR$_{HN}$ counterparts. The findings suggest that present results can play a vital role towards realization of inorganic spintronic devices.

## ACKNOWLEDGMENTS

The authors are thankful to Computational Nanoscience and Technology Laboratory (CNTL), ABV—Indian Institute of Information Technology and Management Gwalior (India) for providing computational facility. One of the authors (HMR) acknowledges the Ministry of Human Resource Development (MHRD), government of India for providing financial support as Teaching Assistantship

# Figure Captions:

**FIG. 1.** Convention of ribbon width and the supercell units ($N_a=9$) for (a) ABNNR$_{HB}$, (b) ABNNR$_{HN}$ and (c) ABNNR$_{HBN}$. The ABNNRs are modeled with periodic boundary conditions along *z*- axis whereas x and y directions are confined.

**FIG. 2.** Half-metallic behavior of 9-ABNNR$_{HB}$ Calculated (a) spindown and (b) spin up electronic band structure, and (c) the total. The spindown (spin-up) bands nearest to the Fermi level in valence and conduction band are indexed with α (α') and β (β') respectively. The Fermi level is indicated by green dotted line whereas spindown (spin-up) states are shown with red (blue) color. Insets; show Transmission coefficient as a function of energy for corresponding ribbon configuration.

**FIG. 3.** Calculated (a) electronic band structure and the total DOS for 9-ABNNR$_{HN}$. Inset; shows Transmission coefficient as a function of energy for 9-ABNNR$_{HN}$. Inset; shows the width dependence of band gap for ABNNR$_{HBN}$. The Fermi level is indicated by green dotted line whereas spindown (spin-up) states are shown with red (blue) color.

**FIG. 4.** Spin degeneracy and insulating nature of 9-ABNNR$_{HB}$. Calculated (a) electronic band structure and (b) the total DOS. Inset (1); shows the width dependence of band gap for ABNNR$_{HBN}$. Inset (2); shows Transmission coefficient as a function of energy for 9-ABNNR$_{HBN}$. The Fermi level is indicated by green dotted line whereas spindown (spin-up) states are shown with red (blue) color.

**FIG. 5.** Calculated Bloch states for spindown α, β, (α-1), (β+1) bands and spin up α' and β' bands of 9-ABNNR$_{HB}$ are presented in the form of (a) Contour and (b) Isosurface. The isovalue is 0.07 (au). The results of α band, for actual ribbon geometry with repeated unit cell which is used for calculations, are presented as an example in the dotted rectangle. The ABNNRs are modeled with periodic boundary conditions along *z*- axis whereas x and y directions are confined.

# Table Captions:

**TABLE I.** Calculated magnetic moment and energy difference ($\Delta E=|E_{LSDA}-E_{LDA}|$) between spin polarized and spin un-polarized states for most stable ABNNR$_{HN}$ with different widths.

**TABLE II.** The overall magnetic and electronic behavior of ABBNRs due to different kinds of edge hydrogenation.

**TABLE III.** Calculated cohesive energies (eV) as a function of ribbon width for all considered ribbon structures.

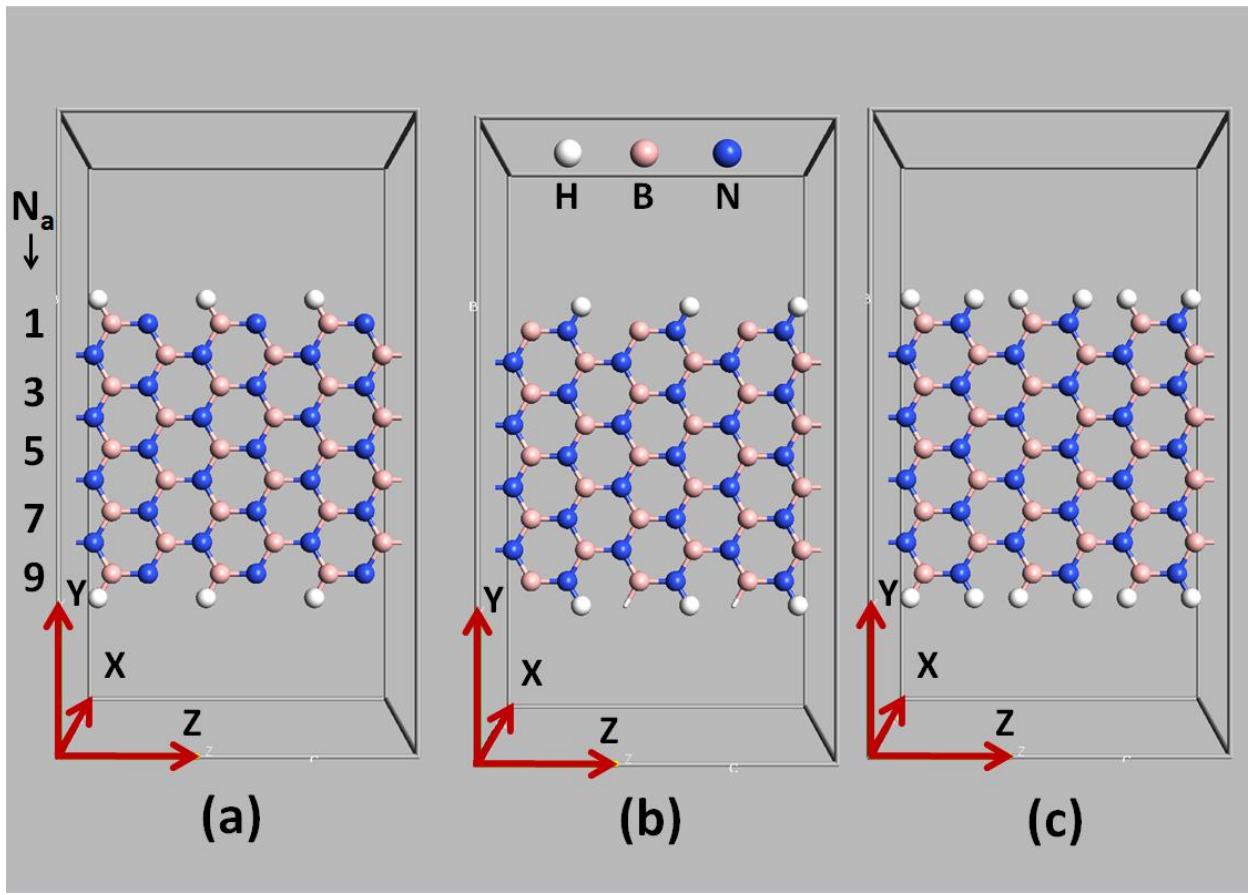

Figure 1    Rai et. al.

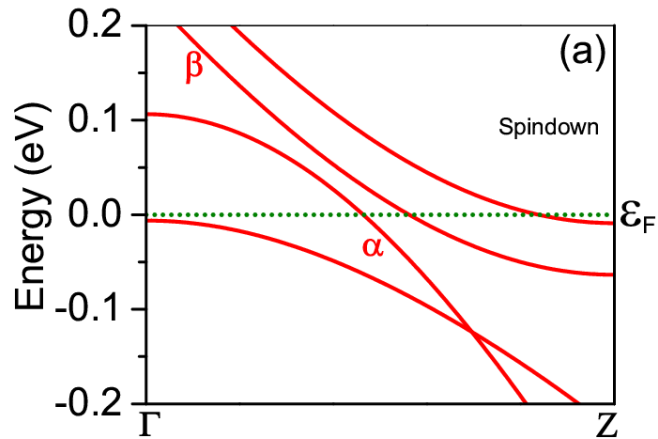

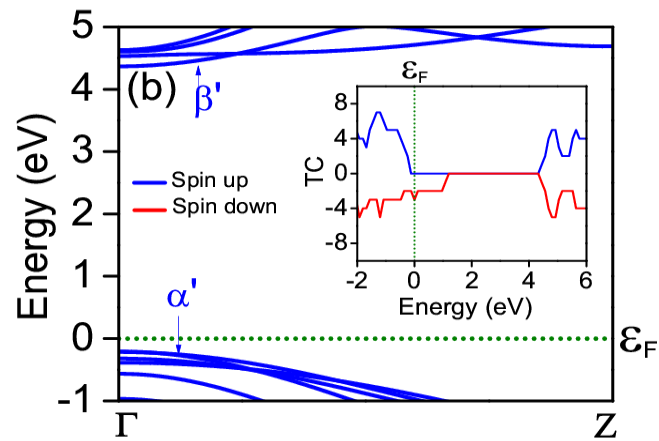

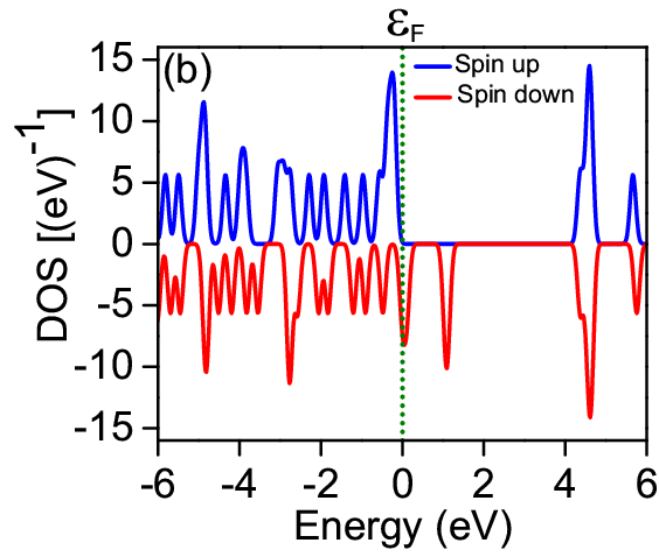

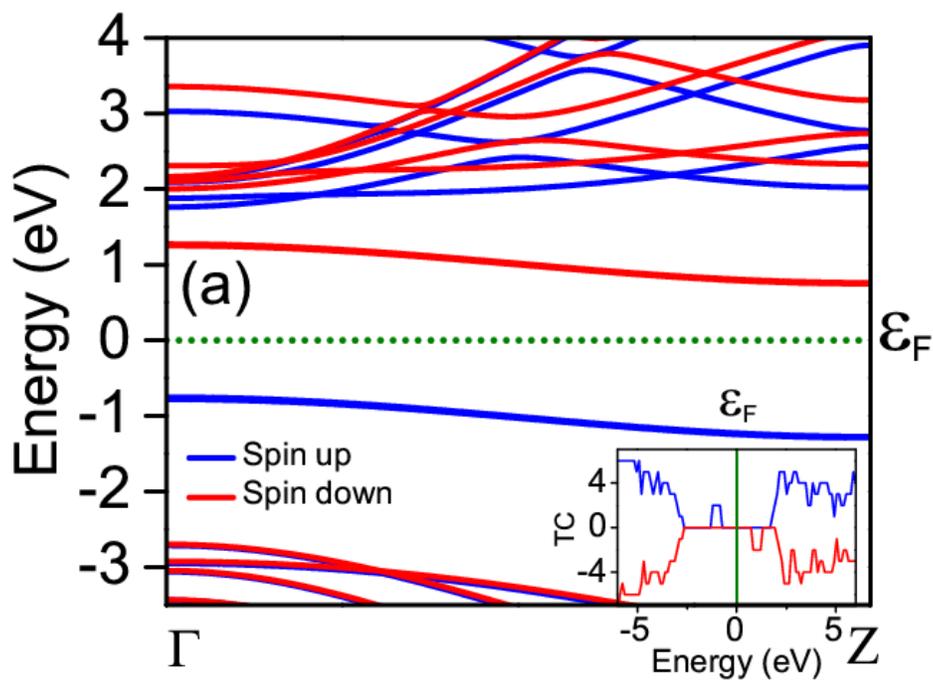
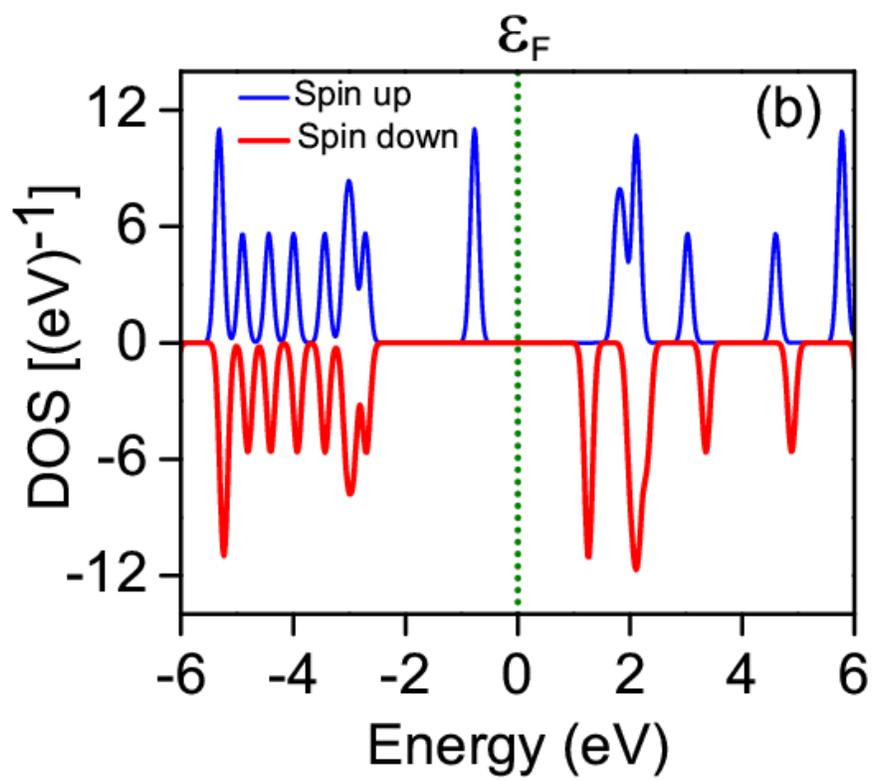

Figure 3         Rai et. al.

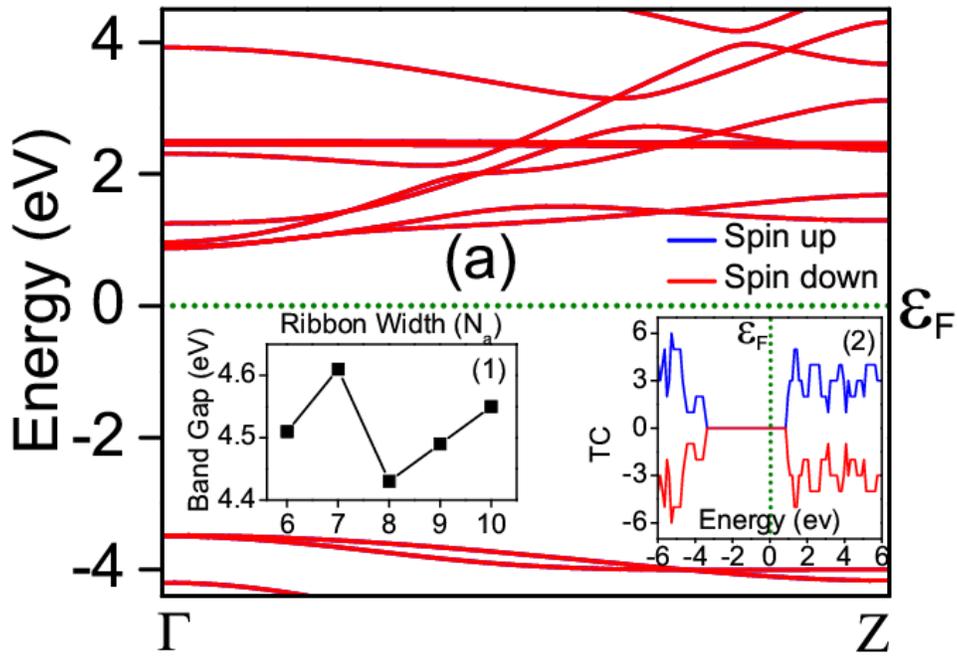

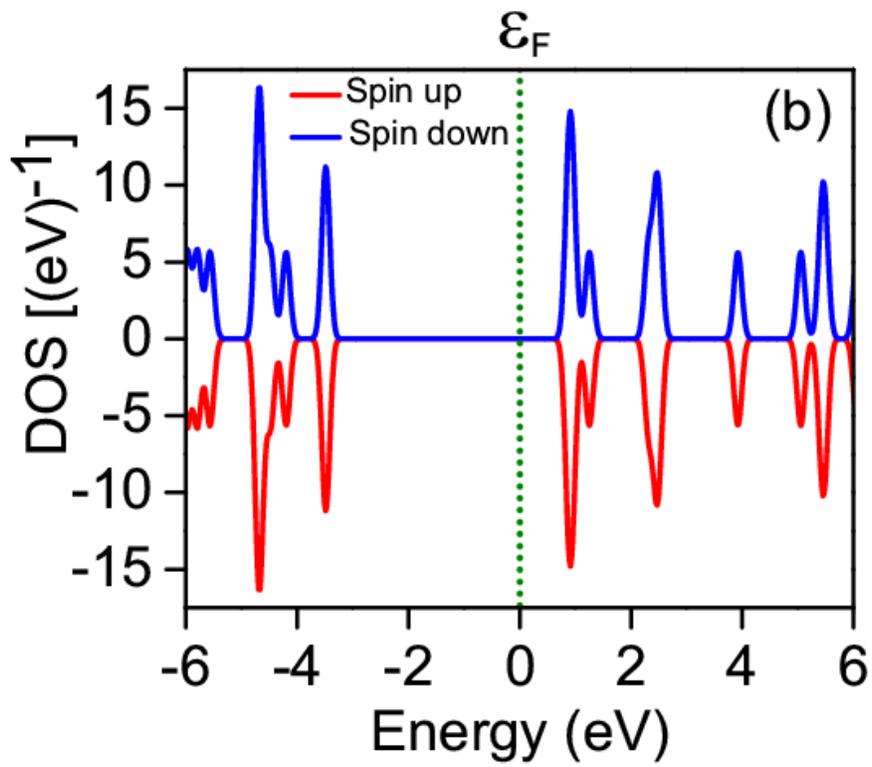

Figure 4     Rai et. al.

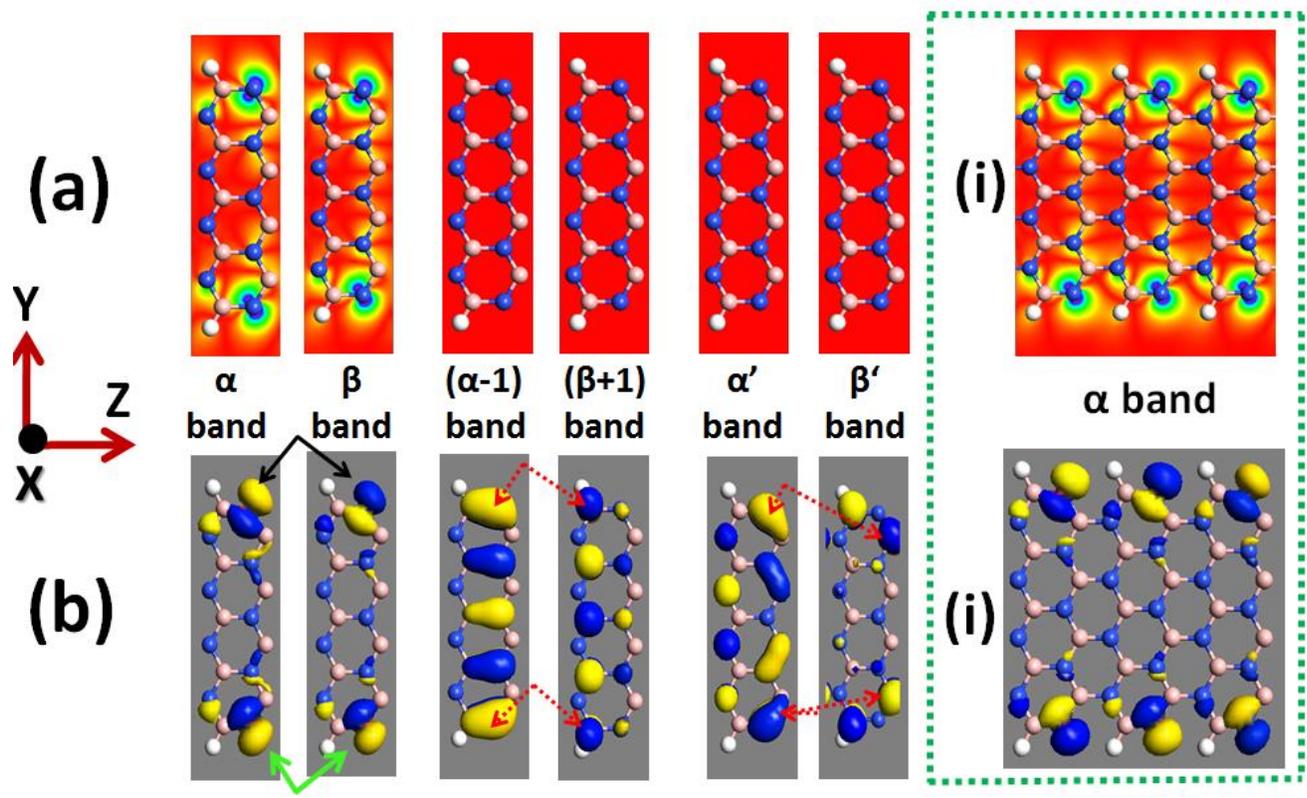

Figure 5    Rai et. al.

TABLE I. Calculated magnetic moment and energy difference ($\Delta E=|E_{LSDA}-E_{LDA}|$) between spin polarized and spin un-polarized states for most stable ABNNR$_{HN}$ with different widths.

|  | $N_a$=6 | $N_a$=7 | $N_a$=8 | $N_a$=9 | $N_a$=10 |
|---|---|---|---|---|---|
| $M$ ($\mu_B$) | 2.01 | 2.02 | 2.01 | 2.02 | 2.02 |
| $\Delta E$ (eV) | 0.55 | 0.55 | 0.56 | 0.55 | 0.55 |

TABLE II. The overall magnetic and electronic behavior of ABBNRs due to different kinds of edge hydrogenation.

| Ribbon Configuration | Avg. $M$ ($\mu_B$) | Electronic state | Band gap (eV) | Magnetic behavior |
|---|---|---|---|---|
| ABNNR$_{HBN}$ | 0 | Insulator | 4.43 to 4.61 | Non- magnetic |
| ABNNR$_{HB}$ | 2.04 | Half-metal (0.26 eV) | 0 | Magnetic |
| ABNNR$_{HN}$ | 2.02 | Semiconductor with a wide band gap | 2.75 to 3.34 | Magnetic |

TABLE III. Calculated cohesive energies (eV) as a function of ribbon width for all considered ribbon structures.

| Ribbon width $N_a$ | $E_c$ (eV) | | |
|---|---|---|---|
|  | ABNNR$_{HBN}$ | ABNNR$_{HB}$ | ABNNR$_{HN}$ |
| 6 | -8.44 | -8.66 | -8.76 |
| 7 | -8.64 | -8.86 | -8.94 |
| 8 | -8.79 | -9.00 | -9.08 |
| 9 | -8.92 | -9.12 | -9.20 |
| 10 | -9.03 | -9.22 | -9.29 |

**Rai et. al.**